\documentstyle[twocolumn,aps,epsfig]{revtex}
\begin{document}
\draft
\title{Critical Effects at 3D Wedge-Wetting}

\author{A.\ O.\ Parry, C.\ Rasc\'{o}n, A.\ J.\ Wood}
\address{Mathematics Department, Imperial College\\
180 Queen's Gate, London SW7 2BZ, United Kingdom}
\date{\today}
\maketitle

\begin{abstract}
We show that continuous filling or wedge-wetting transitions are possible
in 3D wedge-geometries made from (angled) substrates exhibiting
{\it first-order} wetting transitions and develop a comprehensive
fluctuation theory yielding a complete classification of the critical
behaviour. Our fluctuation theory is based on the derivation of a
Ginzburg criterion for filling and also an exact transfer-matrix analysis
of a novel effective Hamiltonian which we propose as a model for
wedge fluctuation effects. The influence of interfacial fluctuations is shown
to be very strong and, in particular, leads to a remarkable
universal divergence of the interfacial roughness $\xi_{\perp}\!\sim\!
(T_F-T)^{-1/4}$ on approaching the filling temperature $T_F$, valid for
all possible types of intermolecular forces.
\end{abstract}

There are two reasons why it is extremely difficult to observe interfacial
fluctuation effects at continuous (critical) wetting transitions in the
laboratory \cite{1}. Firstly, critical wetting is a rather rare phenomenon
for which no examples are known for solid-liquid interfaces and only a limited
number for fluid-fluid interfaces \cite{2,3}. Secondly, the influence of
interfacial fluctuations in three dimensions ($d\!=\!3$) is believed to be
rather small \cite{1}. For example, for systems with long-ranged forces,
the divergence of the wetting layer thickness $\ell$ on approaching the
wetting temperature $T_w$ is mean-field-like, $\ell\!\sim\!(T_w-T)^{-1}$,
and the only predicted effect of fluctuations is to induce an extremely
weak divergence of the width (roughness) $\xi_\perp$ of the
unbinding interface: $\xi_\perp\!\sim\!\sqrt{-\ln(T_w-T)}$. Non-classical
critical exponents and an appreciable interfacial width are only
predicted for systems with strictly short-ranged forces \cite{4}, but
even here the size of the asymptotic critical regime is very small
and beyond the reach of current experimental and simulation
methods \cite{3,5,6}.

The purpose of the present article is to show that these problems
do not arise for continuous (critical) filling or wedge-wetting
transitions \cite{7,8,9}
occurring for fluid adsorption in three-dimensional wedges. First,
we show, contrary to previous statements in the literature \cite{8}, that
critical filling can occur in systems made from walls that
exhibit first-order wetting transitions. Consequently, the observation
of critical filling transitions is a realistic experimental prospect. Second,
we argue that interfacial fluctuations have a strong influence on the character
of the filling transition and, in particular, the interfacial roughness
of the unbinding interface, which is shown to diverge with a universal
critical exponent. The fluctuation theory we develop is based on the
derivation of a Ginzburg criterion for the self-consistency of mean-field (MF)
theory and also an exact transfer matrix analysis of a novel interfacial
Hamiltonian model for wedge wetting which we introduce to account for
the highly anisotropic soft-mode fluctuations. This model leads to a
complete classification of the critical behaviour in $d\!=\!3$ and
predicts some remarkable fluctuation dominated phenomena which we
believe may be tested in the laboratory.

To begin, we recall the basic phenomenology of wedge-wetting and
highlight the mechanism by which critical filling occurs in wedge
geometries even for walls exhibiting first-order wetting transitions.
Consider a wedge (in $d\!=\!3$) formed by the junction of two walls at angles
$\pm\alpha$ to the horizontal (see Fig.\ 1). Axes ($x,y$) are oriented across
and along the wedge respectively. We suppose the wedge is in contact with a
bulk vapour phase at temperature $T$ (less than the bulk critical value $T_c$)
and chemical potential $\mu$. Macroscopic arguments \cite{7,8} dictate that at
bulk coexistence, $\mu\!=\!\mu_{sat}(T)$, the wedge is completely filled by
liquid for all temperatures $T_c\!>\!T\!\geq\!T_F$ where $T_F$ is the filling
temperature satisfying $\Theta(T_{F})\!=\!\alpha$. Here,  $\Theta(T)$ is
the temperature dependent contact angle of a liquid drop on a planar surface.
Thus, filling occurs at a temperature lower than the wetting temperature
$T_w$ and may be viewed as an interfacial unbinding transition (of first-
or second-order) in a system with broken translational invariance. We refer
to any continuous filling transition occurring as $T\!\rightarrow\!T_F$,
$\mu\!\rightarrow\!\mu_{sat}(T_F)$ as critical filling. Also of interest
is the complete filling transition which refers to the continuous
divergence of the adsorption as $\mu\!\rightarrow\!\mu_{sat}(T)$
for $T_c\!>\!T\!\geq\!T_F$ which is known to be characterised by
geometry dependent critical exponents \cite{10}. Here, we focus
exclusively on
critical filling and, in particular, the critical singularities
occurring as $t\!\equiv\!(T_F-T)/T_F\!\rightarrow\!0^{+}$ at bulk
coexistence. The phase transition is associated with the divergence
of four lengthscales (see Fig.\ 1) each characterised by a critical exponent:
the mid-point ($x\!=\!0$) height of the liquid-vapour interface
$\ell_0\!\sim\!t^{-\beta_0}$, the mid-point interfacial roughness
$\xi_{\perp}\!\sim\!t^{-\nu_\perp}$, the lateral extension of the filled region
$\xi_{x}\!\sim\!t^{-\nu_x}$ and the correlation length of the interfacial
fluctuations along the wedge $\xi_{y}\!\sim\!t^{-\nu_y}$. So far, there has
been no discussion of the values of these critical exponents for three
dimensional systems beyond a simple MF calculation for $\ell_0$ \cite{8}.
On the other hand, transfer-matrix studies \cite{9}
in $d\!=\!2$ indicate that fluctuation effects are very strong at
wedge-wetting and lead
to universal critical exponents $\beta_{0}\!=\!\nu_{\perp}\!=\!\nu_{x}\!=\!1$.
This is highly suggestive that fluctuation effects play an
important role in $d\!=\!3$, relevant to experimental studies.

\begin{figure}[t]
\label{first}
\centerline{\epsfig{file=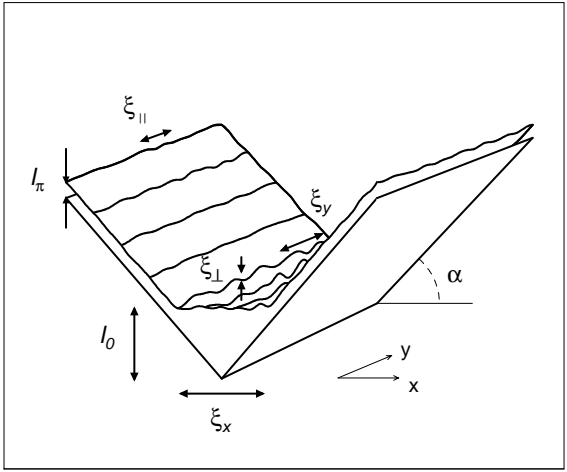,width=10.cm}}
%\vspace*{-1.8cm}
\caption{Schematic illustration of an interface configuration in the
wedge geometry showing the relevant diverging lengthscales at
the filling transition. The planar adsorption $\ell_\pi$
and planar transverse correlation length $\xi_\parallel$ remain finite
at the transition.}
\end{figure}

Previous MF analysis \cite{8} have shown that a suitable starting point for the
study of wedge-wetting in open wedges (small $\alpha$) is the interfacial
model

\begin{equation}
\label{one}
H[\ell]=\int\!\!\int\!dx\,dy
\;\left[\,{\Sigma\over 2}\left(\nabla\ell\right)^2+W(\ell-\alpha|x|)\,\right]
\end{equation}
where $\ell(x,y)$ denotes the local height of the liquid-vapour interface
relative to the horizontal, $\Sigma$ is the
liquid-vapour interfacial tension and $W(\ell)$ is the binding potential
modelling the wetting properties of the wall.
At MF level, this functional is simply minimised to yield an Euler-Lagrange
equation for the $y$-independent equilibrium height profile $\ell(x)$,
$\Sigma\,\ddot\ell\!=\!W'(\ell-\alpha|x|)$,
where the dot and the prime denote differentiation with respect to
$x$ and $\ell$ respectively. This differential equation is solved subject
to the boundary conditions $\dot\ell(0)\!=\!0$ and
$\ell(x)\!-\!\alpha|x|\!\rightarrow\!\ell_\pi$ as $|x|\!\rightarrow\!\infty$.
Here, $\ell_\pi$ denotes the MF planar wetting film thickness
({\it i.e.}, $W'(\ell_\pi)\!=\!0$) and remains microscopic at the
filling transition. Integrating once the equation yields a simple
equation for the midpoint height,
$\Sigma\alpha^{2}/2=W(\ell_0)-W(\ell_\pi)$,
which can be solved graphically \cite{8}. Note that at bulk coexistence,
Young's equation implies $W(\ell_\pi)\!=\!-\Sigma\Theta^{2}/2$ (within
the present small angle approximation) so that the present model
immediately recovers the macroscopic result $\Theta(T_F)\!=\!\alpha$.
Depending on the form of $W(\ell)$ (at $T_F$) the divergence of $\ell_0$
as $T\!\rightarrow\!T_{F}^-$ is first-order or continuous. The condition
for critical filling is that between the global minimum of $W(\ell)$ at
$\ell_\pi$ and the extremum at $\ell\!=\!\infty$ there is no potential barrier.
Thus, walls exhibiting critical wetting necessarily form wedges exhibiting
critical filling. However, for walls exhibiting first-order wetting, the
filling transition is first order or critical depending on whether the
transition temperature $T_F$ is greater or lesser than the spinoidal
temperature $T_s\;(<\!T_w)$ at which the potential barrier in $W(\ell)$
appears. Since the macroscopic condition $\Theta(T_F)\!=\!\alpha$
implies that $T_F$ can be trivially lowered by making the wedge angle
more acute, it follows that walls exhibiting first-order wetting transitions
will, in general, exhibit both types of filling transition
(see Fig.\ 2).
Note that the tricritical value of the wedge angle $\alpha^*$ separating
the loci of first and second-order filling transitions will itself be
rather small for weakly first-order wetting so that the Hamiltonian
(\ref{one}) is still valid. The MF value of height critical
exponent $\beta_0$ for critical filling follows directly from
the equation for $\ell_0$
if we write the asymptotic decay of the binding potential as
$W(\ell)\!\approx\!-\;A\,\ell^{-p}$
where $A$ is a (positive) Hamaker
constant and $p$ depends on the range of the forces. For systems
with short-ranged forces, this decay is exponentially
small. A simple calculation then yields
$\beta_0\!=\!1/p$ (quoted in ref.\ \cite{9} and implicit in reference
\cite{8}) so that, for dispersion forces (corresponding to $p\!=\!2$),
the MF prediction is $\beta_0\!=\!1/2$ whilst for short-ranged forces
$\beta_0\!=\!0(\ln)$. The structure of the MF height profile $\ell(x)$
is particularly simple near critical filling \cite{8} and has crucial
consequences. In essence, the interface is flat ({\it i.e.},
$\ell(x)\!\sim\!\ell_0$) for $|x|\widetilde{<}\,\ell_{0}/\alpha$
whilst for $|x|\widetilde{>}\,\ell_{0}/\alpha$, the height decays
exponentially quickly to its asymptotic planar value $\ell_\pi$ above
the wall. Importantly, the lengthscale controlling this exponential
decay is the wetting correlation length
$\xi_{\parallel}\!\equiv\!\sqrt{\Sigma/W''(\ell_{\pi})}$ which remains
{\it microscopic} at the filling transition. One consequence of this is that the
lateral width of the filled portion of the wedge is trivially identified
as $\xi_{x}\!\sim\!2\ell_{0}/\alpha$ so that $\nu_{x}\!=\!\beta_0$.
More important consequences of the height structure are considered below.

We now turn to the main body of our analysis concerning the nature of
fluctuation effects at critical filling and consider first fluctuations
about the MF profile $\ell(x)$ as measured by the height-height
correlation function $H(x,x';\widetilde{y})\!\equiv\!\langle\,\delta\ell(x,y)\,
\delta\ell(x',y')\,\rangle$ where $\delta\ell(x,y)\!\equiv\!\ell(x,y)
-\langle\,\ell(x,y)\,\rangle$ and $\widetilde{y}\!\equiv\!y'\!-\!y$.
To calculate the correlation function, we first exploit the translational
invariance along the wedge and introduce the structure factor

\begin{equation}
\label{five}
S(x,x';Q)=\int\!d\widetilde{y}\;e^{i\,Q\,\widetilde{y}}\,H(x,x';\widetilde{y}).
\end{equation}

The assumption of MF theory is that fluctuation about $\ell(x)$ are small and
hence a Gaussian expansion of $H[\ell]$ about the minimum suffices to determine
the correlations. This leads to the differential (Ornstein-Zernike) equation

\begin{equation}
\label{six}
\left(-\Sigma\partial^{2}_{x}+\Sigma Q^{2}+
W''(\ell(x)\!-\!\alpha|x|)\right)\;S(x,x';Q)\!=\!\delta(x\!-\!x')
\end{equation}
where we have adsorbed a factor of $k_B T$ into the definitions of $\Sigma$ and
$W(\ell)$. The structure of correlations across the wedge is manifest in the
properties of the zeroth moment $S_0(x,x')\!\equiv\!S(x,x';0)$ which can be
obtained analytically using standard methods. We find

\begin{eqnarray}
\label{seven}
S_0(x,x')=\left(|\dot{\ell}(x)|-\alpha\right)\,
\left(|\dot{\ell}(x')|-\alpha\right)\,\times\,\hspace{.7cm}\\
\nonumber\left\{
\frac{1}{2\alpha W'(\ell_0)}\,+\,\frac{H(xx')}{\Sigma}
\int_{0}^{\min(|x|,|x'|)}\!\!
\frac{dx}{\left(\dot{\ell}(x)-\alpha\right)^2}\right\}
\end{eqnarray}
where $H(x)$ denotes the Heaviside step function ($H(x)\!=\!1$ for
$x\geq 0$, $H(x)\!=\!0$ otherwise). From the properties of the
equilibrium profile $\ell(x)$, it follows that the lengthscale
$\xi_x$ also controls the extent of the correlations across
the wedge. In fact, it can be seen that correlations across the wedge are
very large and also (essentially)
position independent, provided both particles lie within the filled region,
implying that, at fixed $y$, the local height of the filled region
fluctuates coherently. On the other hand, the correlations are totally
negligible if one (or both) particles lie outside the filled region
since their asymptotic decay is controlled by the microscopic
length $\xi_\parallel$. These are important remarks central to the
development of a general fluctuation theory of wedge-wetting.

Turning next to correlations along the wedge, we note that a simple extension
of the above analysis shows that the dominant singular contribution to the
structure factor has a simple Lorentzian form

\begin{equation}
\label{ten}
S(x,x';Q)\;\approx\; \frac{S_0(0,0)}{\,1\,+\,Q^2\,\xi_y^2\,};
\hspace{1cm}|x|,|x'|\widetilde{<}\,\xi_x/2,
\end{equation}
with $S_0(0,0)\!=\!\alpha/2W'(\ell_0)$ which shows a very strong
divergence as $T\!\rightarrow\!T^{-}_F$. The correlation length
along the wedge is identified by $\xi_y\!\approx\!(\Sigma\ell_0/
W'(\ell_0))^{1/2}$.
Substituting for the form of $W(\ell)$, and recalling the divergence of $\ell_0$
at critical filling, leads to the desired MF result $\nu_y\!=\!1/p+1/2$
for the correlation length critical exponent as $T\!\rightarrow\!T_F$
at bulk coexistence. Note that $\xi_y\!\gg\!\xi_x$
so that the fluctuations are highly anisotropic and are totally dominated by
modes parallel to the wedge direction. The final lengthscale that we calculate
within the present MF/Gaussian analysis is the mid-point width $\xi_\perp$
defined by $\xi_\perp^2\!\equiv\!\langle (\ell(0,y)-\ell_0)^2\rangle\!=\!
H(0,0;0)$ which may be obtained from the Fourier inverse of $S(x,x';Q)$.
This leads to the intriguing relation

\begin{equation}
\label{thirteen}
\xi_\perp\;\sim\;\sqrt{\frac{\xi_y}{\,\Sigma\,\ell_0\,}}
\end{equation}
which is one of the central results of this paper. In this way, we are led
to the remarkable prediction that the divergence of $\xi_\perp$ at
critical filling is universal, independent of the range of the intermolecular
forces, and of the form $\xi_\parallel\!\sim\!t^{-1/4}$
which should be observable in experimental and computer simulation studies.
We shall argue below that this result is not affected by fluctuation effects
even when MF theory breaks down.

\begin{figure}[t]
\label{second}
\centerline{\epsfig{file=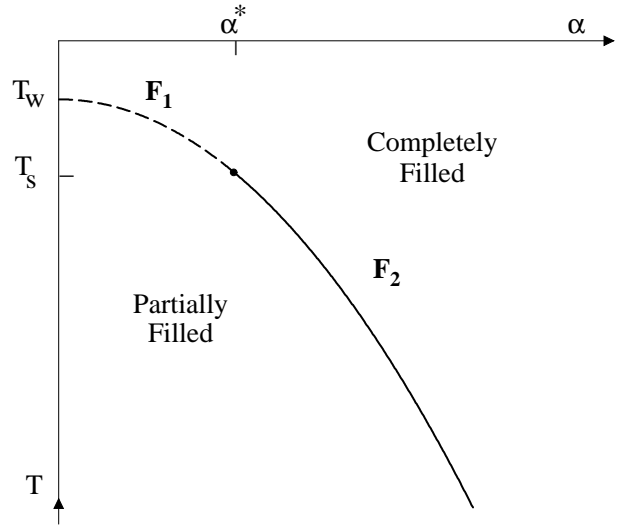,width=10.cm}}
%\vspace*{-1.8cm}
\caption{Schematic surface phase diagram showing 
temperature {\it vs.}
the opening angle $\alpha$ for a system undergoing a first-order
wetting transition at $T_w$ in the planar case ($\alpha\!=\!0$).
The filling transition is only first-order ({\bf F}$_1$) if it takes place
at a temperature above the spinoidal temperature $T_s$ but becomes
second-order ({\bf F}$_2 $) if the filling temperature is less than $T_s$.}
\end{figure}

The first step in the development of a fluctuation theory for filling
transitions is the derivation of a Ginzburg criterion. The MF analysis
presented above should be valid if the fluctuations in the interfacial
height are relatively small. Thus, we require $\xi_\perp/\ell_0\!\ll\!1$
or, equivalently, $t^{1/p-1/4}\!\ll\!1$, implying that MF theory, and the values
of critical exponents quoted above are valid in three-dimensions only for
$p\!<\!4$. For $p\!\geq\!4$, fluctuation effects dominate and we can anticipate
that the roughness $\xi_\perp$ is comparable with the interfacial height
$\ell_0$. One way of approaching this problem is to formulate a renormalization
group theory based on the effective Hamiltonian (\ref{one}). This is an
extremely difficult task and one which we believe is unnecessarily complicated.
In view of the extreme anisotropy of fluctuations at filling transitions
and their coherent nature across the wedge, we propose that the only
fluctuations that are relevant for the asymptotic critical behaviour are those
arising from the thermal excitations of the mid-point height $\ell_0(y)$
along the wedge. More specifically, for a constrained non-planar configuration
for the mid-point distribution $\{\ell_0(y)\}$, we assume that all other
fluctuations are small and, hence, following established methods \cite{11},
may be treated in saddle-point approximation. Thus, we are led to a simpler
wedge Hamiltonian (of reduced dimensionality), $F[\ell_0(y)]\!=\!
\min^{\dagger}H[\ell(x,y)]$
where the dagger denotes the constraint that $\ell(0,y)\!=\!\ell_0(y)
\;\,\forall y$. In this way, we have derived the simpler one-dimensional model
(of three-dimensional filling)

\begin{equation}
\label{sixteen}
F[\ell_0]=\int\!\!dy
\;\left[\,{{\Sigma\ell_0}\over\alpha}\left(\frac{d\ell_0}{dy}\right)^2+
V_F(\ell_0)\,\right]
\end{equation}
where the coefficient of the gradient term is the local height dependent line
tension describing the bending energy of long-wavelength fluctuations along
the wedge and $V_F$ is the effective wedge filling potential which has the
general expansion

\begin{equation}
\label{seventeen}
V_F(\ell)\;=\;\frac{\Sigma(\Theta^2-\alpha^2)}{\alpha}\,\ell +
\frac{A}{(p-1)\alpha}\,\ell^{1-p} + \cdots.
\end{equation}
Note that, in the critical regime,
$(\Theta(T)-\alpha)\!\sim\!t$, so that minimisation of (\ref{seventeen})
identically recovers the MF result for $\ell_0$. For $p\!=\!1$, the second
term in (\ref{seventeen}) is logarithmic whilst for short-ranged forces,
it is exponentially small.

We propose that the effective Hamiltonian (\ref{sixteen}) contains all the
essential physics associated with the asymptotic critical behaviour at filling
transitions. Two checks on this hypothesis are that, in MF and Gaussian
approximation, the new model identically recovers the equation for the mid-point
height and structure factor emerging from the more
complicated model (\ref{one}) in the same approximation. The great advantage of
the new model is, of course, that due to its one-dimensional character, it can
be studied exactly using transfer-matrix techniques. The (normalized)
eigenfunctions $\psi_n(\ell_0)$ and eigenvalues $E_n$ of the
spectrum are found by solving the differential equation (setting
$k_BT\!=\!1$ for convenience)

\begin{equation}
\label{eighteen}
%-\frac{\alpha}{\Sigma\,\ell_0}\,\psi''_n(\ell_0)+
%\frac{3\,\alpha}{2\,\Sigma\,\ell_0^2}\,\psi'_n(\ell_0)+
%V_F(\ell_0)\,\psi_n(\ell_0)\,=\,
%E_n\,\psi_n(\ell_0)
-\frac{\alpha\,\psi''_n(\ell_0)}{\Sigma\,\ell_0}+
\frac{3\,\alpha\,\psi'_n(\ell_0)}{2\,\Sigma\,\ell_0^2}+
V_F(\ell_0)\,\psi_n(\ell_0)\,=\,
E_n\,\psi_n(\ell_0)
\end{equation}
from which the quantities of interest can be calculated.
In particular, the probability distribution for the mid-point height
${\cal P}(\ell_0)\!=\!|\psi_0(\ell_0)|^2$ and the wedge correlation
length $\xi_y\!=\!1/(E_1-E_0)$. The solution of this eigenvalue problem
for the wedge potential (\ref{seventeen}) gives a complete classification of
the critical behaviour at critical filling.
The calculation confirms that MF theory is valid for $p\!<\!4$,
whilst the criticality is fluctuation dominated for $p\!>\!4$ and is
characterised by universal critical exponents
$\beta_0\!=\!\nu_x\!=\!\nu_\perp\!=\!1/4$ and $\nu_y\!=\!3/4$.
These exponents are pertinent to systems critical filling occurring
in systems with short-ranged forces and may be tested in Ising model
simulation studies similar earlier work on critical wetting \cite{5}.
For experimental systems with dispersion forces ($p\!=\!2$), our
predictions are $\beta_0\!=\!\nu_x\!=\!1/2$, $\nu_\perp\!=\!1/4$
and $\nu_y\!=\!1$.

To finish our article, we make two final remarks. Firstly, out of bulk
two-phase coexistence ($\delta\!\mu\!\equiv\!\mu_{sat}(T)-\mu\!>\!0$)
and close to filling, the mid-point height, correlation
lengths and roughness show scaling behaviour.
For example, in the fluctuation-dominated regime, the solution of
(\ref{eighteen}) shows that $\ell_0\!=\!t^{-1/4}\,\Lambda(\delta\!\mu\,t^{-5/4})$
where $\Lambda(\zeta)$ is an appropriate scaling function. Thus, along the
critical filling isotherm ($T\!=\!T_F$, $\delta\!\mu\!\rightarrow\!0$), the height
diverges as $\ell_0\!\sim\!\delta\!\mu^{-1/5}$, which may be easier to observe in
experimental and simulation studies. Secondly, the effective filling model
that we have introduced can also be used to study complete filling occurring
for $T\!>\!T_F$ as $\delta\!\mu\!\rightarrow\!0$. The critical behaviour here
is found to be MF-like ({\it i.e.} $\xi_\perp\!\ll\!\ell_0$) but also universal,
independent of the range of the forces and is consistent with the hypothesis
that the geometry of the wedge determines the critical behaviour for this
transition \cite{10}. Fluctuation effects at this transition are rather less
interesting than for critical filling.

In summary, we have developed a fluctuation theory for critical effects at
three-dimensional wedge-wetting or filling transitions and given a complete
classification of the possible critical behaviour. Fluctuations have been
shown to have a much greater influence in the critical behaviour compared to
that occurring for wetting transitions (in $d\!=\!3$) at planar surfaces
and lead to a universal roughness exponent  $\nu_\perp\!=\!1/4$. We
believe that these predictions are open to experimental verification, in wedge
systems made form substrates exhibiting first-order wetting.

C.R.\ acknowledges economical support from the E.C.\
under contract ERBFMBICT983229. A.J.W.\ acknowledges economical
support from the EPSRC.

\end{document}